# Characterization of Al and Mg alloys from their x-ray emission bands

**Philippe Jonnard, Karine Le Guen**

UPMC Univ Paris 06, CNRS-UMR 7614, Laboratoire de Chimie Physique - Matière et Rayonnement, 11 rue Pierre et Marie Curie, F-75231 Paris cedex 05, France

**Raynald Gauvin, Jean-François Le Berre**

Department of Materials Engineering, McGill University, 3610 University Street, Montréal, Québec, H3A 2B2, Canada

*Abstract*

The valence states of Mg–Al alloys are compared to those of reference materials (pure Mg and Al metals, and intermetallics). Two methods based on x-ray emission spectroscopy are proposed to determine the phases and their proportion: first, by analyzing the Al valence spectra of the Mg-rich alloys and the Mg valence spectra of the Al-rich alloys; second by fitting with a linear combination of the reference spectra the Al spectra of the Al-rich alloys and the Mg spectra of the Mg-rich alloys. This enables us to determine that Al and $Al_3Mg_2$ are present in the 0 – 43.9 wt% Al composition range and Mg and $Al_{12}Mg_{17}$ are present in the 62.5 – 100 wt% Al composition range. In the 43.9 – 62.5%Al range, the alloy is single phase and an underestimation of the Al content of the alloy can be estimated from the comparison of the bandwidth of the alloy spectrum to the bandwidths of the reference spectra.

**Corresponding author** :     Dr. P. Jonnard
Laboratoire de Chimie-Physique, 11 rue Pierre et Marie Curie, F-75231 Paris cedex 05, France
Tel : 33 1 44 27 63 03     Fax : 33 1 44 27 62 26
jonnard@ccr.jussieu.fr

# 1. Introduction

Magnesium alloys are of current interest owing to their low density and high mechanical properties which make them attractive for the transportation industry. Indeed, their use is expected to reduce the weight of vehicles and the fuel consumption, which will reduce air pollution and make the use of automobiles more economical. However, magnesium has a hexagonal closed-packed structure that impedes its ductility at room temperature and this limit its applicability. In order to improve its ductility, new alloys must be developed and an understanding of its band structure with alloying is needed in that regard. The main alloying element of Mg alloys is aluminum. Also, magnesium is an important alloying element of aluminum alloys. Therefore, studying the band structure of Mg–Al alloys is as important in a practical perspective as it is in a fundamental point of view.

We study the emission bands of various dilute Mg alloys and Mg–Al alloys in a large range of composition from x-ray emission spectroscopy (XES) at high spectral resolution. In this work, we consider as a dilute alloy, a metal with less than 6% of another metal. The observed photon energy distribution is the valence density of states weighted by the transition probability and convoluted by the natural width of the core level and the instrumental function. Since the transition probability is almost constant along the valence band and the natural width and the instrumental function are small with respect to the valence bandwidth, the observed photon energy distribution describes the occupied valence density of states. These valence states are sensitive to the chemical environment of the emitting atoms and then the shape of the emission bands changes depending on the neighbors of the Mg or Al emitting atoms.

The emission bands of the alloys are compared to those of references: the pure metals and the two Mg–Al intermetallics. The aim of the study is two fold, to measure the effect of alloying on the band structure of Al and Mg and also to see if it is possible to determine the composition of an alloy directly from its x-ray emission band or by comparing it to those of the references. The effect of a low temperature (400°C) annealing on the emission bands is also reported. X-ray diffraction (XRD) and scanning electron microscopy (SEM) are used to complement the XES results.

## 2. Materials and methods

### 2.1. Sample preparation

For a given alloy, pure rods of the elements were cut and weighed precisely for obtaining a given composition. They were put in a separate graphite cubicle that was molten in a Lindberg furnace under argon atmosphere at a temperature of 750 ºC. The liquid alloy was cast in a copper mold that was water cooled.

The compositions of the samples were determined by inductively coupled argon plasma-atomic emission spectrometry (ICAP-AES). The aimed composition is the actual one to within ±2%. The main impurities in the samples are found to be Fe (0.01-0.1%), Zn (0.04-0.06%) and Si (0.007-0.03%). The other impurities are less than 0.003%. All the given concentrations are weight concentrations unless otherwise specified.

All the alloys were cut and polished down to the 0.1 µm level for their observation in the scanning electron microscope and their XES analysis. To see if the casting microstructure has an effect on the Mg Kβ and Al Kβ emission bands, some alloys were annealed for 6 days at 400°C (673 K) in an argon atmosphere.

### 2.2. X-ray diffraction

The XRD analysis was performed at a wavelength of 0.154 nm using the Cu Kα radiation. XRD was used to determine the phases present in the alloys.

## 2.3. Scanning electron microscopy

The SEM analysis was performed on the Hitachi SN-3000 using backscattered electrons and x-ray microanalysis with an Oxford EDS detector. Quantitative x-ray microanalysis was performed using a new intensity ratio method (Gauvin et al., 2006a). This new method is based on the ratio of the net x-ray intensity of the Mg Kα line to the sum of the net intensity of the Mg Kα and the Al Kα lines. This ratio of net intensities is then related to the Mg composition of the alloy with a calibration curve computed with the Monte Carlo Win X-ray program (Gauvin et al., 2006b). The Al composition is them obtained by difference. The energy of the incident electrons was set to 25 keV. This energy was mainly chosen as a compromise between imaging with backscattered electrons and microanalysis. Only the overall composition of the alloys has been measured by performing x-ray microanalysis at low magnification (x50).

## 2.4. X-ray emission spectroscopy

The XES experiments are performed in home built high resolution x-ray spectrometer (Bonnelle *et al.*, 1994) working in the WDS mode. The sample is placed under vacuum and excited by an electron beam. Following the ionization of the atoms present in the sample, x-rays are emitted (Azaroff, 1974; Bonnelle, 1987), then dispersed and detected in a high-resolution bent crystal spectrometer. We have measured the Al Kβ and Mg Kβ emission bands coming from the aluminum and magnesium atoms present in the samples. They correspond to the 3p-1s transition (1s ionization threshold at 1560 and 1300 eV for Al and Mg, respectively) and describe the occupied valence states having the Al 3p or Mg 3p character.

The advantage of studying these emissions is that the transitions imply valence electrons that participate directly to the chemical bond (Vergand *et al.*, 1989; Jonnard *et al.,* 1998; Jarrige *et al.*, 2002; Jonnard *et al.*, 2004; Kashiwakura & Nakai, 2004). Then, the electronic transitions that take place from the valence band to an inner shell are very sensitive to the chemical environment of the emitting atom. In the present work, we compare the spectra of the various studied alloys to those of the pure metals (Al and Mg) and their main intermetallics ($Al_3Mg_2$ and $Al_{12}Mg_{17}$). Two other advantages come from the involvement of an inner shell in the emission process. First, this makes XES characteristics of the emitting atom, *i.e.* if one observes the Mg (Al) emission, there will be no interference from the emission coming from the Al (Mg) atoms. Second, this makes XES a local probe, *i.e.* it is sensitive to the first neighbors around the emitting atoms but almost not to the long range order.

The size of the electron beam is about 1 $cm^2$. The energy of the incident electrons was set to 7 keV in order to probe the bulk of the samples. The analyzed thickness is estimated to be about 500 nm. The electron density impinging on the sample is sufficiently low (<1 $mA/cm^2$) to avoid any evolution of the sample under the electron beam. This is confirmed by checking that the shape and intensity of the studied emission remain unchanged throughout the measurement. For each spectrum, a linear background is subtracted. The maximum of each spectrum is also normalized to unity.

## 3. Results

### 3.1. Pure metals and intermetallics

We present in Figure 1 the Mg Kβ and Al Kβ emission bands of pure metals and intermetallics. They describe, respectively, the distribution of the occupied Mg 3p and Al 3p states within the valence band. For each spectrum a discontinuity appears toward the high photon energies that marks the position of the Fermi level. It can be seen that the emission band of each compound has its own shape and width. This will enables us to determine in the alloy samples the contribution of each phase. It is observed that the widths of the Mg emission band with decreasing Al content. The same effect, but less pronounced, is observed for the Al emission band, i.e. the width decreases with the Al content. The presented spectra are in agreement with spectra obtained previously on similar apparatus (Neddermeyer, 1972, 1973; Nemoshkalenko, 1972).

Toward the high photon energies, it is observed for the pure metals that the background intensity is lower than that toward the low energies. This is due to the self-absorption effect : indeed above the Fermi level exists a strong density of unoccupied states, leading to a strong jump of the absorption coefficient. Thus, when the energy of the emitted photon is higher than the absorption edge (about 1560 and 1300 eV for Al and Mg respectively), a large part of the intensity is absorbed within the sample. This explains why the background is lower at high photon energy than at low photon energy (Bonnelle, 1987). The jump in the region of the Mg (Al) 1s absorption edge is stronger when the Mg (Al) content is higher within the sample.

### 3.2. Dilute Mg alloys

The Mg K$\beta$ and Al K$\beta$ emission bands of various Mg dilute alloys are presented in the figure 2. For MgLi (6% Li) the Mg K$\beta$ emission is very close to that of pure Mg. The same is true for the spectrum of the MgAlZn (3% Al, 1% Zn) alloy except in the region 1290-1295 eV where some extra intensity is observed. This is probably due to an oxidation of the sample because this photon energy range corresponds to the maximum of the Mg K$\beta$ band from MgO (figure 2a). The MgO spectrum is obtained from a single crystal cleaved along the (100) plane. This is also confirmed by observing the chemical state of the Al atoms in the superficial zone of this sample, i.e. by using low energy electrons of 2.5 keV probing the first 70 nm of the sample. In this case the Al K$\beta$ emissions shows the feature characteristic of oxidized Al. This is not observed for the other samples.

Concerning the Al Kβ emission bands of MgAl (3% Al) and MgAlZn (1% Al, 1% Zn), a strong decrease of the bandwidth is observed (Figure 2b) as expected from the evolution of the bandwidth with the Al content in the reference spectra. In these cases, the quality of the spectra is poorer because of the low counting statistics due to the small number of Al atoms in these samples.

### 3.3. AlMg alloys

Figure 3 shows the Mg Kβ and Al Kβ emission bands of the Mg–Al alloys compared to those of the pure metals. As expected from the analysis of the reference spectra (Figure 1) the bandwidth decreases with the Al content of the alloy. It is observed that the band shape is close to that of the pure Mg (Al) when the Mg (Al) concentration is high in the alloy. When the Al concentration decreases (Fig. 3c), the shape of the top of the Al Kβ emission becomes similar to that of the intermetallics. The self-absorption effect of the Mg (Al) band is stronger when the Mg (Al) content is higher.

There is no significant evolution of their spectrum shape with the annealing despite a refinement of their grain size after the annealing. This is true for the other alloy compositions.

Three samples have been analyzed by SEM: $Al_{85}Mg_{15}$, $Al_{50}Mg_{50}$ and $Al_{30}Mg_{70}$. The corresponding images are presented in Figure 4. It is observed that the Al-rich and Mg-rich alloys are two-phase whereas the $Al_{50}Mg_{50}$ sample is made of a single phase. An energy dispersive spectrometry analysis was performed to determine the mean weight composition of the samples. For the studied samples, the overall composition is in agreement with the one expected from the preparation. Formally, the quantitative analysis method that was used is valid for homogeneous materials only and since figure 4 shows that many alloys have two phases, the overall compositions are an approximation in these cases and the agreement could be fortuitous.

# 4. Discussion

## 4.1. Dilute Mg alloys

The shape of the Mg Kβ band of the dilute alloys is very close that of pure Mg (Figure 2a). This shows that the majority of Mg atoms are not disturbed by the foreign atoms (Li, Al or Zn) due to the low concentration. Because in the dilute Mg alloy the Al atoms are mostly surrounded by Mg atoms (and not by Al atoms as in Al metal) and because the shape of an emission band is sensitive to the chemical state of the emitting atom, the shape of the Al Kβ emission of Al within the Mg alloy is different from that of Al within Al metal. In this case, the Al 3p states strongly interact with the Mg 3p states leading to an Al Kβ spectrum of the dilute alloy looking like the Mg Kβ spectrum of Mg metal.

It is not possible to determine the composition of the dilute alloys from their valence band because of the lack of sensitivity of the bandwidth for small concentration changes. Then, for the Mg emission, the shape of the band of the alloy is very similar to that of pure Mg, and for the Al emission band the statistics is too poor to distinguish too close concentrations.

## 4.2. AlMg alloys

### 4.2.1. Effect of annealing

The 400°C annealing has no particular effect on the emission bands of the studied alloys (not shown). This means that the local arrangement around the Mg or Al does not evolve upon this annealing, even if an evolution of the long range order occur by the increase of the grain size of the alloys. There is no significant evolution of the short range order that could lead to some modification of the shape of the emission bands. In fact the alloys keep their electronic structure.

### *4.2.2. Analysis from the bandwidth of the emission band*

Figure 5 presents the evolution of the bandwidths of the Mg and Al emission from the Mg–Al alloys as a function of the Al mass concentration within the samples, in comparison to those of the pure metals and intermetallics. The bandwidth is determined within ±0.1 eV. The bandwidths of the dilute Mg alloys are also indicated. Our measurements of the Al Kβ bandwidth follow the same trend than those of Tanaka and Matsumoto (Tanaka & Matsumoto, 1974) but are systematically lower. This difference is probably due to a small oxidation of the samples (at least of the pure Al) studied by these authors and to the use of a spectrometer having a lower spectral resolution than ours.

In the concentration range between the two intermetallics (43.9–62.5% Al), it is observed that the Mg and Al bandwidths of the alloys follow those of the references, but are lower than these latter by some tenths of eV. Outside this latter concentration range, the bandwidths are found to be almost constant around about 5.6 eV for the Mg emission toward the high Al content and 3.6 eV for the Al emission toward the low Al content, values close to those of the Mg band of $Al_3Mg_2$ and of the Al band of $Al_{12}Mg_{17}$, respectively (see the horizontal lines in Figure 5). If the bandwidth could be directly used to determine the alloy concentration, then we would deduce, for the $Al_{85}Mg_{15}$ sample for example, that this alloy has a composition close to that of the $Al_3Mg_2$ intermetallics (i.e. 62.5 wt% Al) from its Mg emission bandwidth. Of course, this is not the case. In fact, this means that this sample is made of two phases : the $Al_3Mg_2$ intermetallics and the pure Al. Indeed, when looking at the Mg emission band, there can be no interference with the Al spectrum of pure Al (Cf. Section 2.4) so the Mg alloy bandwidth is that $Al_3Mg_2$. Making the same reasoning for the other samples, one can deduce that for concentrations from 0 to that of $Al_{12}Mg_{17}$ (i.e. 43.9% Al) there exist both phases : Mg and $Al_{12}Mg_{17}$; for concentrations from that of $Al_3Mg_2$ (i.e. 62.5% Al) to 1 there exist both phases : Al and $Al_3Mg_2$. This is confirmed by the XRD measurements. As an example, we display in Figure 6 the diffraction patterns of an Al-rich and a Mg-rich alloy.

From the Al and Mg weight fractions of the samples and knowing the involved intermetallic compound and pure metal, we calculate for the samples having an Al fraction outside the 43.9–62.5% range, the molecular fractions of the intermetallics and the pure metal, *i.e.* the relative contribution of the Al and intermetallics molecules ($Al_3Mg_2$ or $Al_{12}Mg_{17}$), as well as the mass fraction of each phase. They are indicated in Table I for the concerned samples. From both (intermetallics and metal) calculated molecular fractions, by taking into account the stoichiometry, we calculate these contributions normalized with respect to the number of Mg or Al atoms. Then, a spectrum of the Al (Mg) band for the Al (Mg)-rich alloys is computed as a weighted sum of the Al (Mg) and $Al_3Mg_2$ ($Al_{12}Mg_{17}$) spectra, the weights being the previously determined contributions. The width of the computed spectra are determined and compared in Table I to the experimental ones of the corresponding alloys. When there exist more than one sample for a given concentration, the experimental values are the mean of the bandwidth for each sample. It can be seen in Table I that the calculated and experimental bandwidths are in relatively good agreement. However, due to the lack of precision on the experimental bandwidths, it is not possible to accurately discriminate between the various alloys.

Inside the 43.9–62.5% Al concentration range, the Al and Mg bandwidths vary with the composition. This can be explained in two ways:

- the two intermetallics at least are present and the observed width results from the composition of the bandwidths of the various corresponding spectra;
- there is one compound and the observed width is characteristic of the studied composition.

Of course, as indicated by the SEM image (Figure 4) the second case takes place with our samples. Thus to determine the composition of the alloys one should use Figure 5 as an abacus. But because the alloy curves is shifted downward with respect to the reference curves, this would lead to an underestimation of the alloy concentration of about 10%.

### *4.2.3. Analysis of the fitted emission bands*

In order to determine the phases within the alloys, we propose to fit the alloy spectra by a weighted sum (linear combination) of the reference spectra. This is now routinely done for the study of complex materials (Kurmaev *et al.*, 1995; Iwami *et al.* 1997; Galakhov, 2002; Jonnard *et al.*; 2005; Maury *et al.*, 2006a; Maury *et al.*, 2006b; Salou *et al.*, 2008).

As an example, we present in Figure 7 the fit of the Mg K$\beta$ band of a magnesium-rich alloy ($Al_{30}Mg_{70}$) and that of the Al K$\beta$ band of an aluminium-rich alloy ($Al_{90}Mg_{10}$). We also present on the same figure, the fits for an alloy ($Al_{50}Mg_{50}$) having an Al content intermediate between those of $Al_3Mg_2$ and $Al_{12}Mg_{17}$. The fits are made with the minimal number of possible references. We collect in Table II for all the alloy samples and for both emission bands the molecular fractions of the various references. When there is more than one analysis for a sample, the mean values are indicated in the table. From the different contributions, the mass composition deduced from one emission is compared to the one deduced from the other emission and to the nominal composition. The Mg emission of the $Al_{90}Mg_{10}$ alloy is not fitted because of the poor statistics of the spectrum.

Inside the concentration range of the intermetallics (43.9–62.5% Al), the alloy is single phase. Thus, it is not possible to fit the spectra of these alloys because the references (intermetallics and pure metals) are absent. Outside the concentration range of the intermetallics (0–43.9% Al and 62.5–100% Al), as expected from the analysis of the bandwidths (Table I), the Mg K$\beta$ band of the Mg-rich alloys can be fitted by a mix of pure Mg and $Al_{12}Mg_{17}$ spectra, and the Al K$\beta$ band of the Al-rich alloys can be fitted by a mix of pure Al and $Al_3Mg_2$ spectra. The calculated mass fraction of the samples deduced from the relative contributions of the metal and the intermetallics is in agreement with the nominal composition within about 5%. This is why the same fit of the Al band is found for the three alloys: $Al_{80}Mg_{20}$, $Al_{85}Mg_{15}$ and $Al_{90}Mg_{10}$.

From the previous analysis based on the bandwidths and from these fits, one would expect to find only the $Al_{12}Mg_{17}$ intermetallics contribution for the fit of the Al K$\beta$ band of the Mg-rich alloys and only the $Al_3Mg_2$ contribution for the fit of the Mg K$\beta$ band of the Al-rich alloys. In fact this is not true for two reasons :

- even if each reference spectrum has its own particular shape, these shapes are somewhat sufficiently close so that it is possible to introduce a small fraction of another spectrum in a fit without changing the overall agreement;
- when the concentration of Al (Mg) is low, the counting statistics of the Al (Mg) band is poorer than that of the Mg (Al) band, so that it is possible to find more than one contribution to the Al (Mg) emission band.

Thus, in the concentration range outside those of the intermetallics, it is preferable to only consider relevant the fit of the Mg K$\beta$ band for the Mg-rich alloys and the fit of the Al K$\beta$ band for the Al-rich alloys.

The Mg and Al $L_{2,3}$ (3sd – 2p transition) emission bands could have been used for this kind of study. Their shape is very different from that of the K$\beta$ bands because they do not involve the same valence states. For both K and L emission bands the same kind of evolution is observed as with the Al fraction : progressive change of the shape from one composition to another and increase of the bandwidth. Thus, the L emission bands could be treated in the same way we have done for the K emission bands. In fact the L bands have already been used to study the electronic structure of various Al and Mg alloys (Das Gupta & Wood, 1955; Gale & Trotter, 1956; Appleton & Curry, 1965; Nemoshkalenko, 1972; Neddermeyer, 1972, 1973;

Szasz *et al.*, 1988) as well as quasi-crystals involving Al and Mg atoms (Fournée *et al.*, 1999; Belin-Ferré *et al.*, 2000). However, the L emission bands of the light metals are in the ultra-soft x-ray range and require a grazing incidence x-ray spectrometer in order to obtain the high spectral resolution necessary to be sensitive to the chemical state.

Other emissions from the Mg and Al spectra could also be envisaged, such as the K$\alpha$ atomic lines. They do not involve valence electrons so their shape is not sensitive to the chemical states of the emitting atoms but the position of their maximum is sensitive to it (Day, 1963). However the shifts are quite small, a few tenths of eV. Nevertheless these lines are very intense and so could be useful to study the electronic structure of dilute alloys (Jonnard *et al.*, 2003).

## 5. Conclusions

When the overall concentration of the AlMg alloys is between 0–43.9% Al and 62.5–100% Al, they are made of Al and $Al_3Mg_2$ for the Al-rich alloys and Mg and $Al_{12}Mg_{17}$ for the Mg-rich alloys. It is possible to determine the composition of an Al-rich alloy either from the measure of the bandwith of its Mg K$\beta$ emission or from the fit of its Al K$\beta$ emission with the Al and $Al_3Mg_2$ reference spectra. It is possible to determine the composition of a Mg-rich alloy either from the measure of the bandwith of its Al K$\beta$ emission or from the fit of its Mg K$\beta$ emission with the Mg and $Al_{12}Mg_{17}$ reference spectra. The composition can be determined within a 5% uncertainty. Inside the concentration range of the intermetallics (43.9–62.5% Al), the AlMg alloys are made of one phase. In this case an Al underestimated composition can be obtained from the examination of the bandwidths.

# ACKNOWLEDGEMENTS

The National Science and Engineering Research Council of Canada is acknowledged for financial support concerning this work. Pierre Vermette, from McGill University, is acknowledged for having prepared the alloys.

# TABLES

Table I : For the alloys having an Al mass fraction outside the range of the intermetallics (0-43.9% Al - 62.5-100% Al), molecular fractions (%) and mass fractions (wt%) of the intermetallics and the pure metal and full width at half maximum (ΔE) of the emission band calculated from these contributions compared to the experimental width of the alloy band. If a sample is made of x molecules of the phase X and y molecules of the phase Y, then the molecular fraction of phase X is x/(x+y) and that of phase Y is y/(x+y). The mass fraction of phase X is $xM_X/(xM_X+yM_Y)$ and that of phase Y is $yM_Y/(xM_X+yM_Y)$, where $M_X$ and $M_Y$ are respectively the molar weight of X and Y species. In our case X and Y can be Al, Mg, $Al_3Mg_2$ or $Al_{12}Mg_{17}$.

| Sample | % Mg %wgt | % $Al_{12}Mg_{17}$ %wgt | % $Al_3Mg_2$ %wgt | % Al %wgt | ΔE (eV) cal | ΔE (eV) exp |
|---|---|---|---|---|---|---|
| $Al_{15}Mg_{85}$ | 98.3 65.9 | 1.7 34.1 | - | - | 3.3 | 3.5 ± 0.1 |
| $Al_{30}Mg_{70}$ | 93.4 31.7 | 6.6 68.9 | - | - | 3.7 | 3.8 ± 0.1 |
| $Al_{65}Mg_{35}$ | - | - | 74.3 93.3 | 25.7 6.7 | 5.0 | 4.7 ± 0.1 |
| $Al_{70}Mg_{30}$ | - | - | 45.4 79.9 | 54.6 20.1 | 5.1 | 5.1 ± 0.1 |
| $Al_{80}Mg_{20}$ | - | - | 19.2 53.3 | 80.8 46.7 | 5.2 | 5.0 ± 0.1 |
| $Al_{85}Mg_{15}$ | - | - | 12.2 40.0 | 87.8 60.0 | 5.2 | 5.2 ± 0.1 |
| $Al_{90}Mg_{10}$ | - | - | 7.0 26.6 | 93.0 73.4 | 5.3 | 5.2 ± 0.1 |

Table II : Molecular fractions (%) of the pure metals and the intermetallics as determined from the fit of the Al and Mg Kβ emission bands of the alloys. For each fit, the weight composition of the sample is deduced taking only the two phases considered in the fit. The relevant fits are indicated in bold style. For instance, for an alloy fitted by a weighted sum with a c1 contribution of the Mg spectrum and a c2 contribution of the $Al_{12}Mg_{17}$ spectrum (with c1 + c2 = 1), the molecular fraction of the Mg phase is c1/(c1+c2/17) whereas that of the $Al_{12}Mg_{17}$ phase is (c2/17)/(c1+c2/17).

| Samples | Mg Kβ fit | | | composition | Al Kβ fit | | | composition |
|---|---|---|---|---|---|---|---|---|
| | % Mg | % $Al_{12}Mg_{17}$ | % $Al_3Mg_2$ | | % $Al_3Mg_2$ | % $Al_{12}Mg_{17}$ | % Al | |
| $Al_{15}Mg_{85}$ | **95.7** | **4.3** | **-** | **$Al_{25}Mg_{75}$** | - | 37.9 | 62.1 | $Al_{47}Mg_{53}$ |
| $Al_{30}Mg_{70}$ | **91.5** | **8.5** | **-** | **$Al_{32}Mg_{68}$** | - | 45.7 | 54.3 | $Al_{46}Mg_{54}$ |
| $Al_{65}Mg_{35}$ | 12.0 | - | 88.0 | $Al_{61}Mg_{39}$ | **85.9** | **-** | **14.1** | **$Al_{64}Mg_{36}$** |
| $Al_{70}Mg_{30}$ | - | - | 100 | $Al_{63}Mg_{37}$ | **85.9** | **-** | **14.1** | **$Al_{64}Mg_{36}$** |
| $Al_{80}Mg_{20}$ | 13.3 | - | 86.7 | $Al_{61}Mg_{39}$ | **19.0** | **-** | **81.0** | **$Al_{80}Mg_{20}$** |
| $Al_{85}Mg_{15}$ | 6.0 | - | 94.0 | $Al_{62}Mg_{38}$ | **19.0** | **-** | **81.0** | **$Al_{80}Mg_{20}$** |
| $Al_{90}Mg_{10}$ | - | - | - | | **19.0** | **-** | **81.0** | **$Al_{80}Mg_{20}$** |

**FIGURE LEGENDS**

**Figure 1** : Mg Kβ (a) and Al Kβ (b) emission bands of pure metals and intermetallics.

**Figure 2** : Mg Kβ (a) and Al Kβ (b) emission bands of various dilute alloys compared to those of pure metals and MgO. MgLi : 6% Li; MgAlZn : 1 or 3% Al and 1% Zn; MgAl : 3% Al.

**Figure 3** : Mg Kβ (a,b) and Al Kβ (c,d) emission bands the AlMg alloys compared to those of the pure metals.

**Figure 4** : Scanning electron microscope image of the $Al_{85}Mg_{15}$ (a), $Al_{50}Mg_{50}$ (b) and $Al_{30}Mg_{70}$ (c) alloys taken with backscattered electrons.

**Figure 5** : Full width at half maximum of the Mg Kβ and Al Kβ emission bands as of function of the Al mass fraction for the intermetallics and the pure metals (references) and for dilute Mg and AlMg alloys. The lines are to guide the eyes. The upper horizontal line is the mean of the Mg Kβ bandwidths of alloys between $Al_{70}Mg_{30}$ and $Al_{90}Mg_{10}$. The lower horizontal line is the mean of the AlKβ bandwidths of alloys between $Al_{15}Mg_{85}$ and $Al_{30}Mg_{70}$.

**Figure 6** : XRD patterns of the $Al_{85}Mg_{15}$ (a) and $Al_{15}Mg_{85}$ (b) alloys compared to the positions of the peaks of the reference materials: Al, Mg, $Al_3Mg_2$ and $Al_{12}Mg_{17}$.

**Figure 7** : Fits of the Mg Kβ (a,c) and Al Kβ (b,d) emission bands of $Al_{30}Mg_{70}$ (a), $Al_{90}Mg_{10}$ (b) and $Al_{50}Mg_{50}$ (c,d), by a weighted sum of the reference spectra.

**FIGURES**

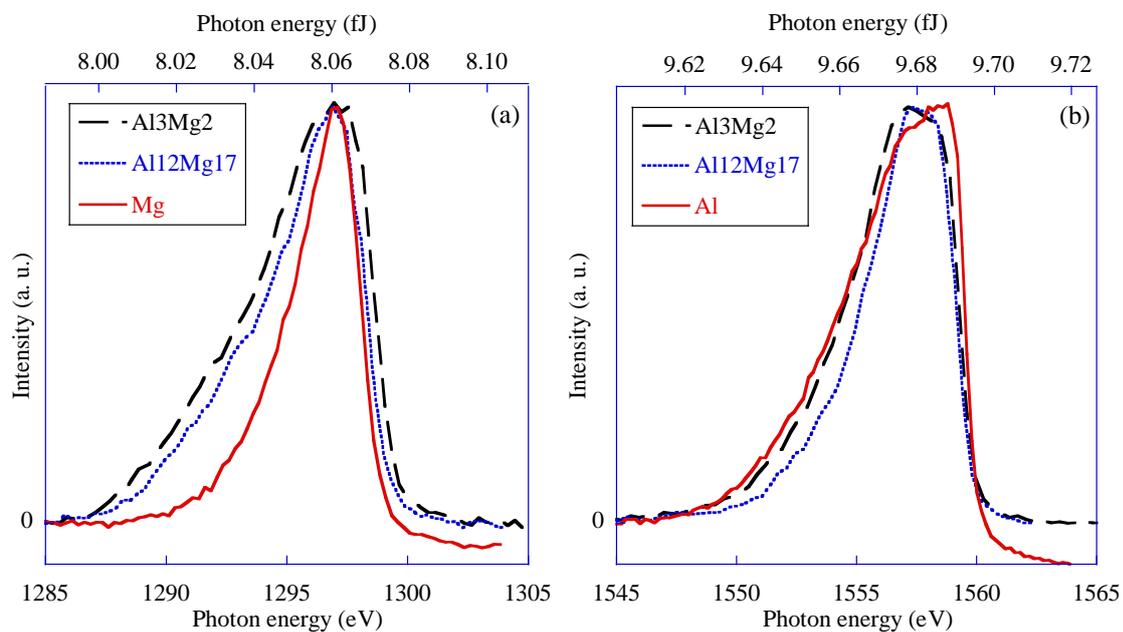

Figure 1

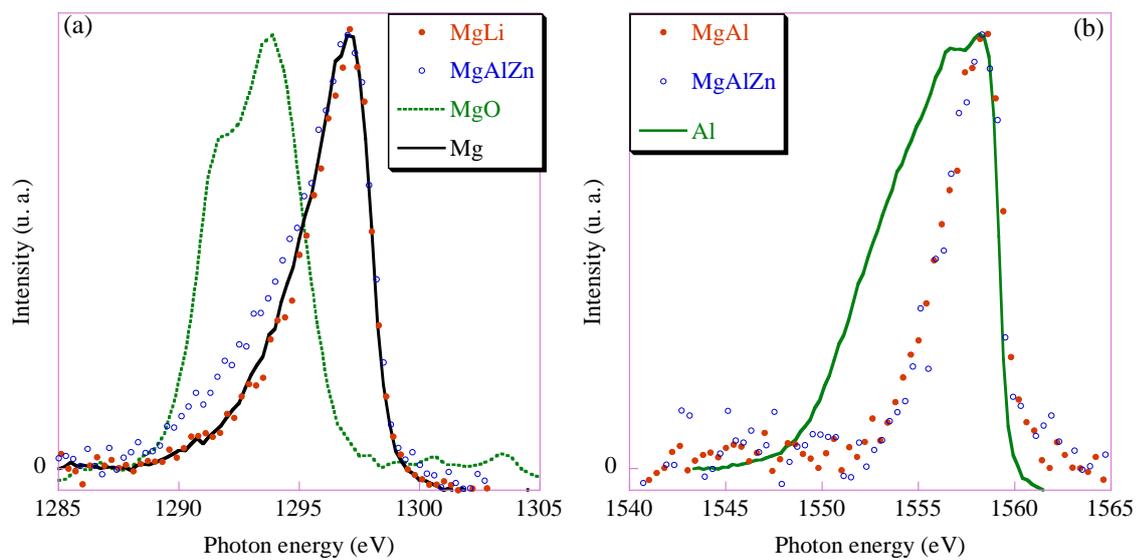

Figure 2

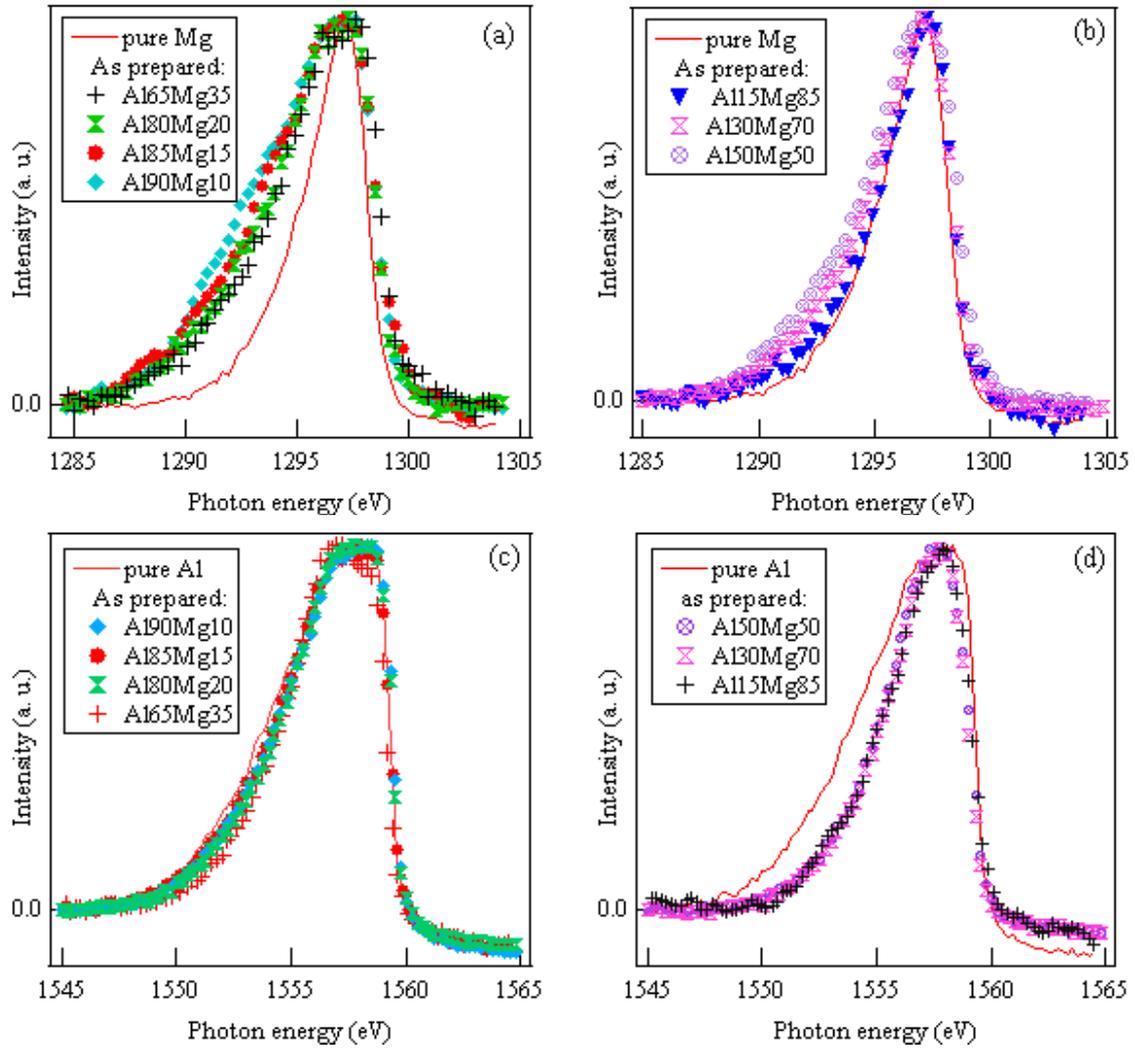

Figure 3

Figure 4

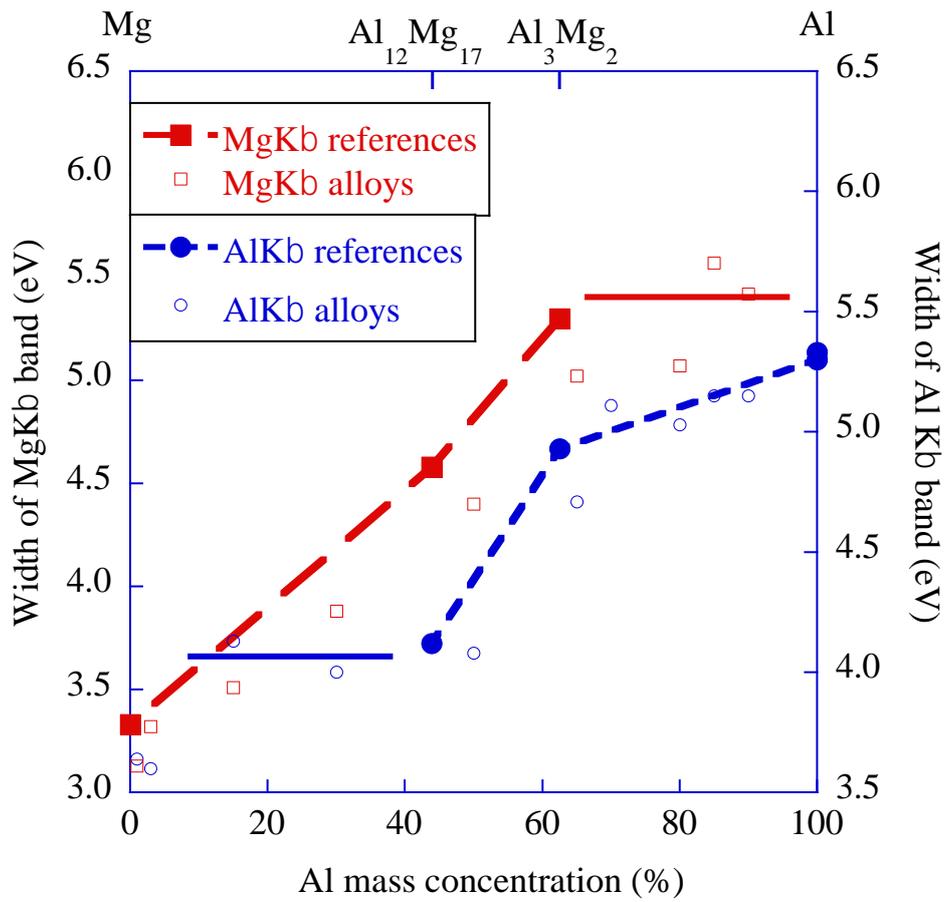

Figure 5

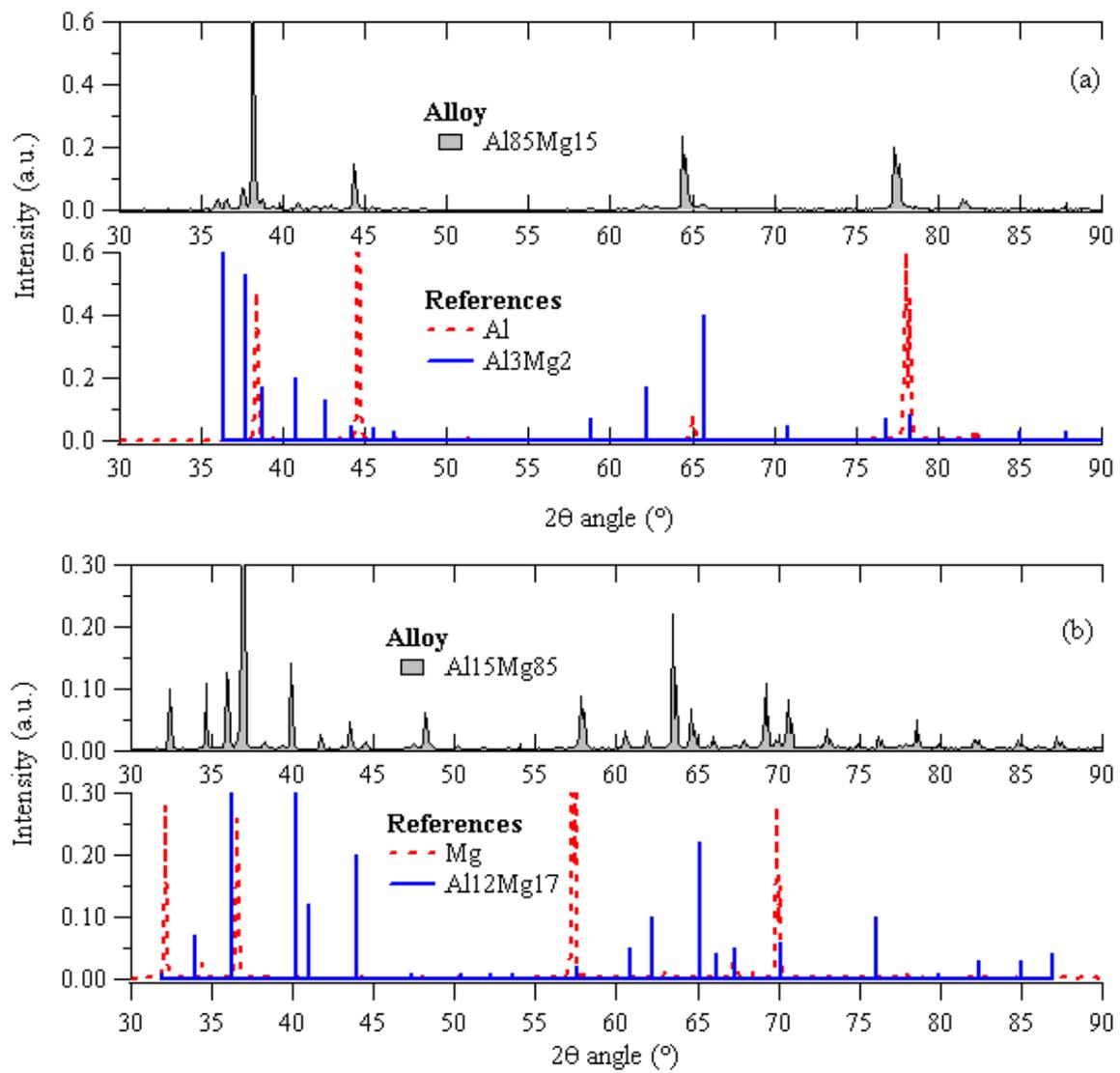

Figure 6

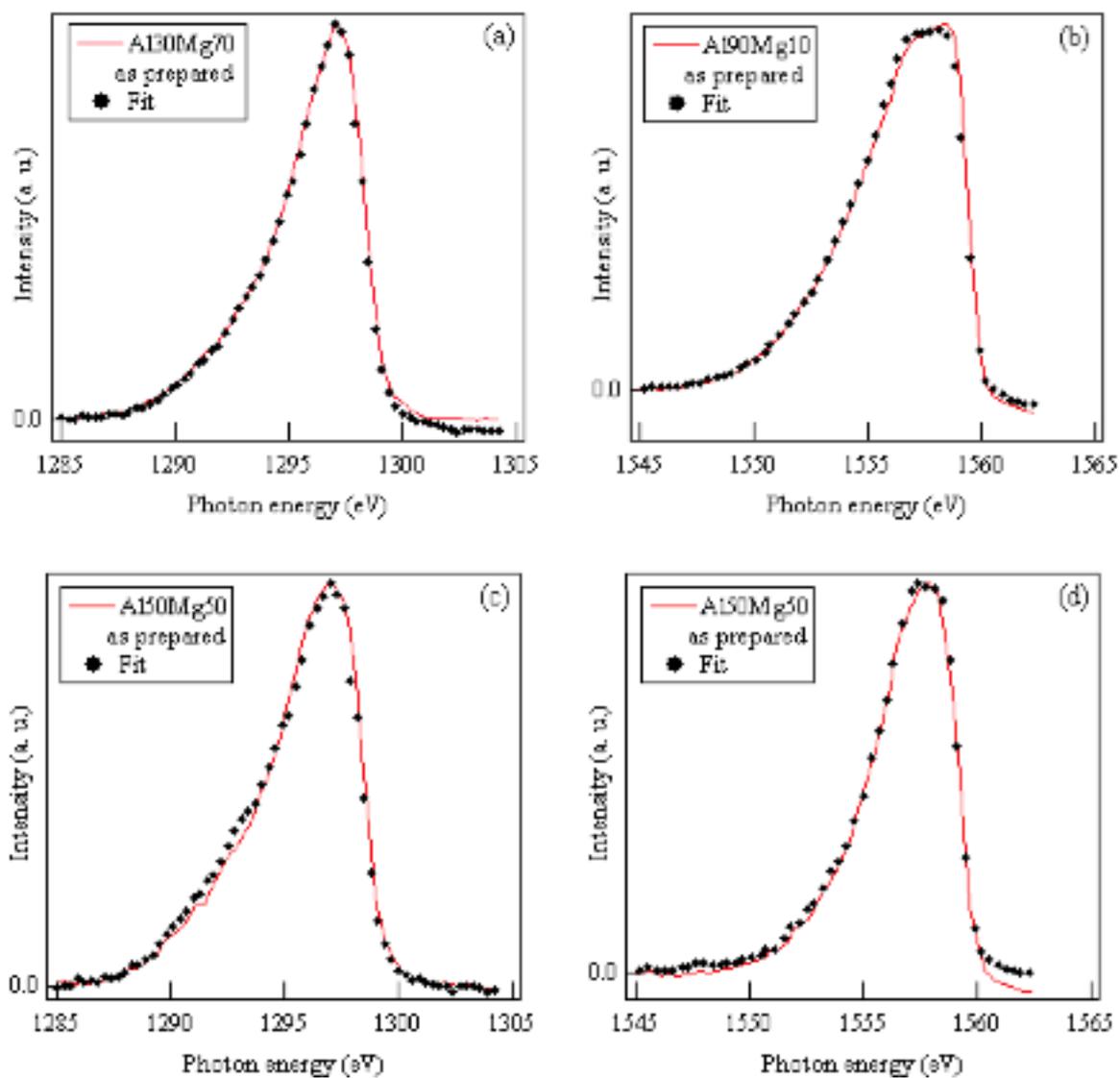

Figure 7